# Establishing Atomic Coherence in Twisted Oxide Membranes Containing Volatile Elements

*Young-Hoon Kim [†,\*], Reza Ghanbari [†], Min-Hyoung Jung, Young-Min Kim, Ruijuan Xu\*, Miaofang Chi\**

Dr. Y.-H. Kim, Prof. M. Chi
Center for Nanophase Materials Sciences (CNMS)
Oak Ridge National Laboratory (ORNL), Oak Ridge, TN 37831, USA
E-mail: kimy6@ornl.gov, chim@ornl.gov

R. Ghanbari, Prof. R. Xu
Department of Materials Science and Engineering
North Carolina State University, Raleigh, NC 27695, USA
E-mail: rxu22@ncsu.edu

Dr. M.-H. Jung, Prof. Y.-M. Kim
Department of Energy Science
Sungkyunkwan University (SKKU), Suwon 16416, Republic of Korea

Prof. Y.-M. Kim
Center for 2D Quantum Heterostructures
Institute for Basic Science (IBS), Suwon 16419, Republic of Korea

Prof. M. Chi
Department of Mechanical Engineering and Materials Science
Duke University, Durham, North Carolina 27708, USA

[†]Y-H. Kim and R. Ghanbari contributed equally to this work.

**Abstract**

Twisted oxide membranes represent a promising platform for exploring moiré physics and emergent quantum phenomena. However, the presence of amorphous interfacial "dead" layers in conventional oxide heterostructures impedes coherent coupling and suppresses moiré-induced interactions. While high-temperature thermal treatments can facilitate interfacial bonding, additional care is needed for materials containing volatile elements, where elevated temperatures may cause elemental loss. This study demonstrates the realization of atomically coherent, chemically bonded interface in twisted $NaNbO_3$ heterostructures through controlled oxygen-annealing treatment. Atomic-resolution imaging and spectroscopy reveal ordered perovskite registry accompanied by systematic lattice contraction and modified electronic structure at the twisted interface, providing signatures of chemical reconstruction rather than physical adhesion. This reconstructed interface mediates highly asymmetric strain propagation in which the bottom membrane remains nearly relaxed while the top membrane accommodates substantial shear strain, thereby establishing a strain gradient that enables long-range electromechanical coupling throughout the twisted oxide membranes. By resolving the nature of the reconstructed interface, these findings establish a robust pathway for achieving coherent and strain-tunable oxide moiré superlattices, opening pathways to engineer emergent ferroic and electronic functionalities.

## 1. Introduction

The emergence of twistronics has sparked tremendous research interest in condensed matter physics and materials science, leading to the discovery of a wealth of intriguing moiré phenomena.[1,2] By introducing controlled angular misalignment between two crystalline layers, moiré superlattices generate long-range modulation of local atomic registry that fundamentally influences materials properties by modifying its electronic band structure and interlayer interactions. To date, most research on moiré engineering have focused on two-dimensional (2D) van der Waals (vdW) materials, where weak vdW coupling between atomically thin layers enable rotational stacking and formation of moiré superlattices.[3] This structural freedom has led to a series of remarkable discoveries, including flat-band superconductivity,[4–6] correlated insulating states,[7,8] emergent magnetism,[9–11] and engineered band structures.[12–14]

Extending twistronics beyond weakly bonded vdW systems to epitaxial complex oxides opens a vastly richer landscape that exploits the strong lattice-charge-spin-orbit coupling intrinsic to transition metal oxides. Recent advances in fabricating freestanding oxide membranes have established oxides as new functional building blocks for constructing moiré heterostructures.[15–18] Freed from substrate constraints through selective wet etching of a sacrificial layer or mechanical exfoliation,[19,20] these membranes can be stacked or twisted at arbitrary orientations, akin to 2D vdW materials, enabling three-dimensional structural tunability and access to coupled charge, orbital, spin, and lattice degrees of freedom beyond conventional heteroepitaxial systems.[16,21–25] Twisted oxide heterostructures exhibit robust ionic-covalent interlayer bonding, giving rise to long-range strain modulation and emergent moiré phenomena including flexoelectricity-driven polar vortices, charge modulation, flat electronic bands, and unconventional magnetism.[17,26–29] Their mechanical flexibility further enables deliberate strain engineering, allowing controlled modulation of ferroic order and the creation of moiré superlattices with programmable functionalities.

Despite the promise, realizing controlled moiré structures in twisted oxide heterostructures present far greater challenges than in vdW materials, as their strong covalent bonding requires the establishment of interfacial chemical connections rather than weak physical adhesion.[29–32] Moreover, individual freestanding oxide membranes often develop a thin amorphous surface layer during release or transfer,[33] originating from surface hydration, residual polymer contamination, or partial crystalline-to-amorphous transition at the 2D limit. This amorphous layer inhibits the formation of strong interfacial bonding at the twisted interface, necessitating additional processing steps to remove or recrystallize the surface and recover atomic coherence.[34–36] While recent work has demonstrated clean interfaces between cubic $SrTiO_3$ membranes and hexagonal sapphire substrates through high-temperature $CO_2$ laser annealing ($\geq$ 1000 ºC),[29] such approache may introduce stoichiometric deviations in oxides that contain volatile elements such as sodium (Na), potassium (K), lithium (Li), lead (Pb), and bismuth (Bi), which are key constituents in many ferroelectric systems. The high volatility of these elements leads to non-stoichiometry, vacancy formation, and increased leakage current at elevated temperatures. Therefore, developing a robust alternative processing strategy is essential for achieving atomically clean interfaces in ferroelectrics containing volatile elements while preserving their structural integrity and functional properties.

In this work, we demonstrate the successful elimination of interfacial contaminants and amorphous layer and the establishment of atomically coherent interfaces in twisted $NaNbO_3$ bilayers (t-NNO) through systematically controlled post-annealing under an oxygen atmosphere. Advanced microscopic and spectroscopic analyses suggests that the t-NNO exhibits ordered –Nb–Na–Nb– atomic sequences and lattice modifications at the interface, confirming the formation of chemical bonding and electronic coupling, rather than weak vdW contact. This complete interfacial bonding generates asymmetric strain fields that propagate across the top membrane, whereas the bottom membrane is nearly relaxed. This study provides a fundamental understanding of how atomic-scale interfacial bonding governs unique strain propagation in twisted oxide bilayers, providing insights toward the design of adaptive, lead-free electronic, energy-harvesting, and quantum functional devices based on oxide heterostructures.

## 2. Results and Discussion
### 2.1. Interfacial challenges in twisted oxide membranes

Epitaxial heterostructures consisting of 6.7 nm-thick NaNbO$_3$ thin films and 20 nm-thick La$_{0.7}$Sr$_{0.3}$MnO$_3$ buffer layers were synthesized on (001)-oriented single-crystalline SrTiO$_3$ substrates via pulsed laser deposition (PLD).[37] Following growth, the buffer layer was selectively etched using a mixed diluted solution of hydrochloric acid and potassium iodide to release freestanding NaNbO$_3$ membranes.[38] The released membrane was then transferred onto a target substrate (e.g., Si grids) using a poly(methyl methacrylate) (PMMA) support layer, which was subsequently removed through dissolution in acetone. To evaluate the crystalline quality and phase structure, X-ray diffraction analysis was performed, revealing high crystalline and a single-phase perovskite structure of the released membranes (**Figure S1**). A second NaNbO$_3$ membrane was prepared via identical procedures and stacked onto the first membrane with a relatively angular offset to form twisted NaNbO$_3$ bilayers (t-NNO), as schematically illustrated in **Figure 1a**.

Plan-view annular dark-field (ADF) imaging demonstrates the formation of a moiré superlattice of as-prepared t-NNO arising from controlled crystallographic misalignment, with a relative twist angle of approximate 47° (**Figure 1b**). Despite the successful twist assembly observed in plan-view imaging, cross-sectional ADF imaging reveals an interfacial region (**Figure 1c**) exhibiting structural characteristics that deviate substantially from those of the adjacent crystalline layers. While both the top and bottom membranes preserve their crystalline periodicity, each approximately 4 nm thick with a consistent pseudo-cubic lattice spacing of $d_{(001)pc}$ = 0.392 nm, an unintended layer with reduced crystalline order was observed at the interface. Low-magnification ADF imaging further corroborates the existence of interfacial roughness and discontinuity extending laterally over 100 nm (**Figure S2a**). Atomic-resolution analysis confirms a clear reduction in ADF intensity within the interlayer region (**Figure S2b-d**). Additional cross-sectional analyses along the zone axes of both membranes confirm high crystalline ordering and structural integrity, although an interfacial layer of approximately 1-3 nm thickness is still observed at the interface (**Figure S3**). The attenuation of ADF intensity indicates the non-crystalline nature of the interface and highlights the inherent difficulty in forming structurally coherent and chemically bonded interfaces between twisted oxide membranes. Such structural discontinuities represent a key challenge, as amorphous interfaces may preclude the realization of emergent functionalities anticipated in twisted oxide membrane systems.[36]

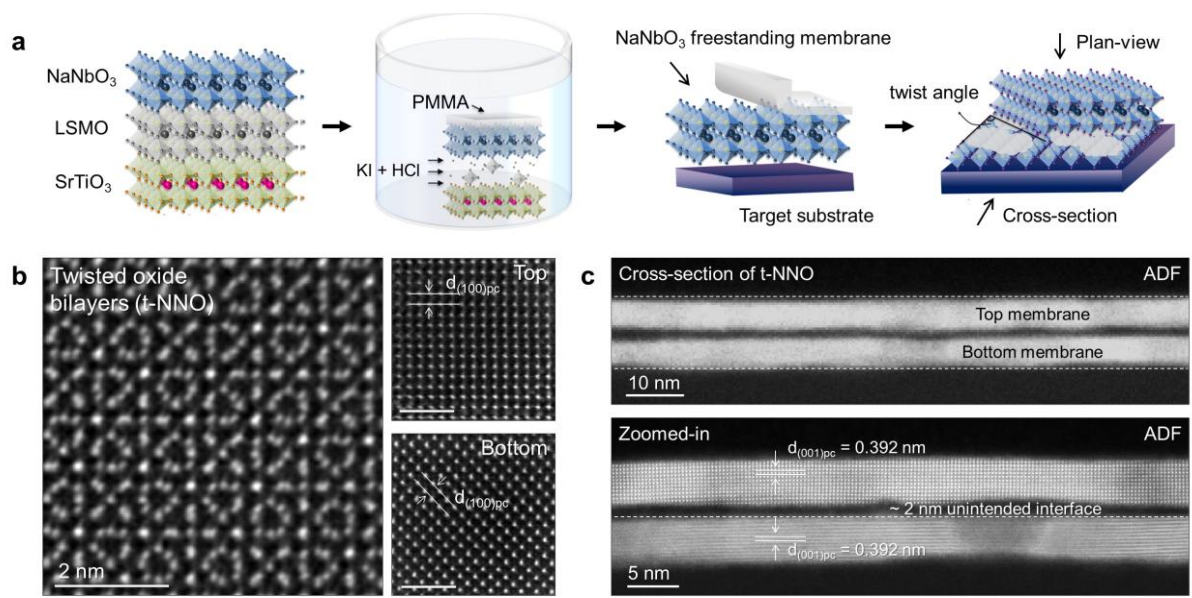

**Figure 1. Fabrication and structural analysis of twisted freestanding membranes.** a) Schematic illustration of the assembly process for twisted NaNbO₃ membranes (t-NNO), showing selective etching of the buffer layer to release freestanding membranes and stacking with angular offset. b) Plan-view annular dark-field (ADF) imaging showing the moiré superlattice of the as-stacked t-NNO with twist angle of ~47°. c) Cross-sectional ADF images revealing a structurally disordered interfacial layer (~2 nm) between the two crystalline NaNbO₃ membranes, indicating interfacial roughness and discontinuity.

To fully address this issue, it is essential to identify the origin of the amorphous interfacial layer. Energy-dispersive X-ray spectroscopy (EDX) was therefore employed to elucidate the chemical characteristics of the unintended interlayer. Elemental mapping reveals the uniform distribution of the main constituents, Na (green), Nb (purple), and O (cyan), across the crystalline membrane regions (**Figure 2a**). However, these constituent elements exhibit substantial depletion at the interface, accompanied by high carbon enrichment. Corresponding line profile analysis (**Figure 2b**) shows that the concentrations of Na, Nb, and O at the interface were reduced by approximately 50% relative to those in the crystalline regions. The spatial extent of this depletion corresponds to the amorphous layer thickness determined via cross-sectional ADF imaging. As a result, the carbon signal is sharply localized at the interface, indicating that carbonaceous contamination constitutes the primary component of the disordered region. These findings suggest that the amorphous interlayer originates from residue PMMA contaminants that were not completely removed during acetone cleaning.[29]

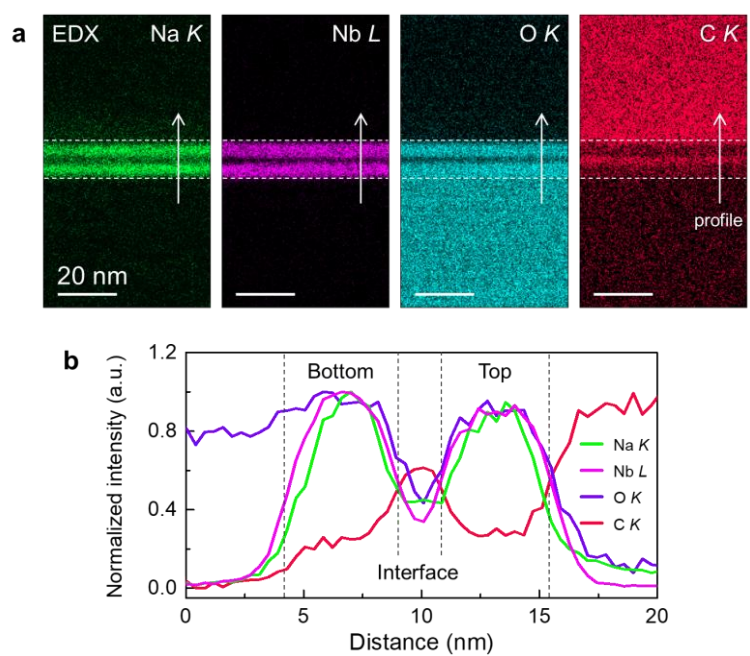

**Figure 2. Chemical analysis of the amorphous interlayer in twisted membranes.** a) Elemental mapping of Na *K* (green), Nb *L* (purple), O *K* (cyan) and C *K* (red). b) Corresponding compositional profile across t-NNO membranes. The interfacial region exhibits a depletion of Na, Nb, and O signals, accompanied by the prominent presence by carbon enrichment.

## 2.2. Atomically flat and coherent interface in twisted oxide membranes

To mitigate such contamination and achieve clean interfaces, we applied a post-annealing treatment in a tube furnace under flowing oxygen. The oxygen-rich environment is expected to promote oxidation that facilitates the removal of carbonaceous species while potentially contributing to film densification and reduced oxygen vacancies. To determine the optimum annealing condition, systematic treatments were conducted on twisted membranes at varying annealing temperature, followed by AFM and optical microscopy to track changes in surface topography and membrane color contrast. AFM analysis reveals a continuous decrease in root-mean-square (RMS) roughness with increasing annealing temperature, from 960 pm in the as-transferred t-NNO (**Figure S4a**) to 186 pm after annealing at 660 °C (**Figure S4f**). In addition, the emergence of two sets of step terraces on membranes annealed at 660 °C for two hours suggests effective interfacial reconstruction between the two membranes. Notably, extending the annealing duration results in minimal improvement, with no significant reduction in RMS roughness (**Figure S4c, d**). Further increasing the annealing temperature to 800 °C results in a pronounced color change in the membranes (**Figure S5**), indicative of severe sodium loss due to its volatility at elevated temperatures. Based on these observations, an optimised annealing condition of 660 °C for two hours under continuous $O_2$ flow (14 mL/min) was selected, consistent with the $NaNbO_3$ growth temperature and significantly lower than the typical annealing temperatures used in previous studies.

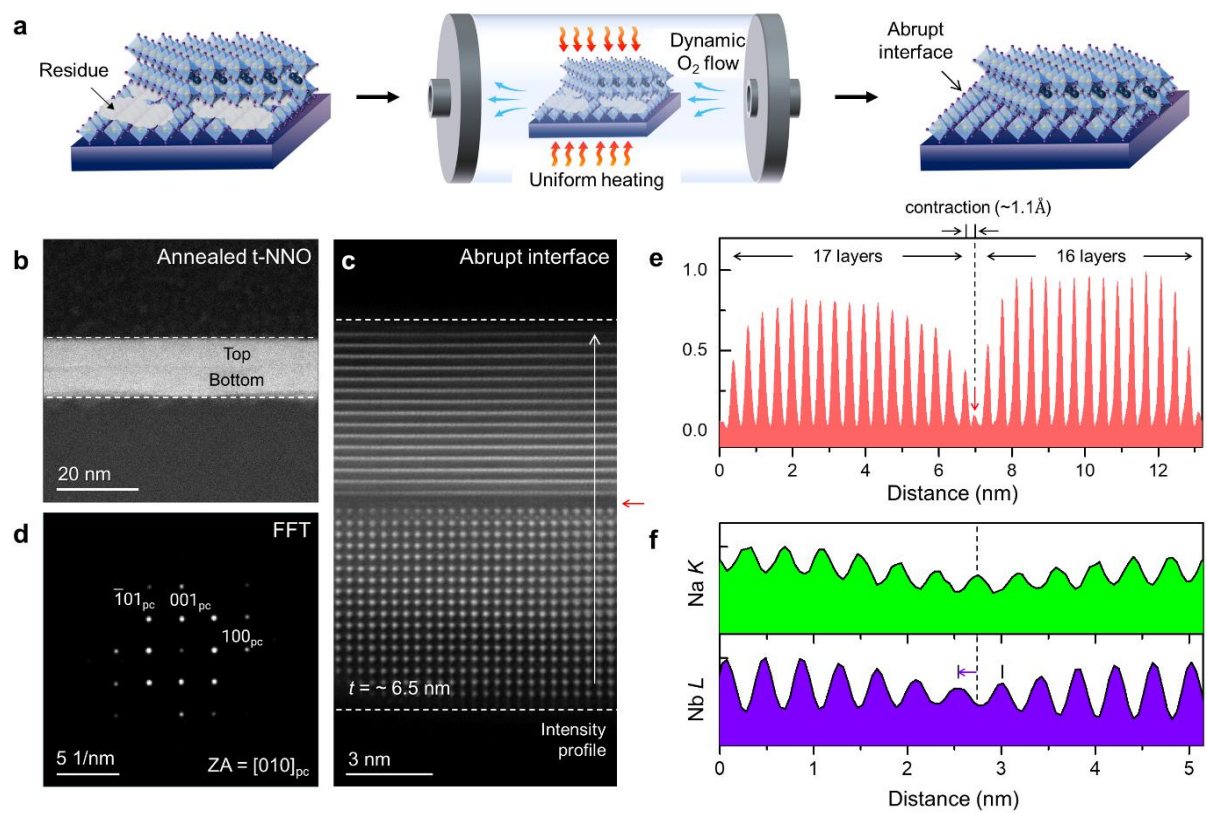

**Figure 3. Formation of an atomically coherent interface in annealed t-NNO**. a) Schematic illustration of the post-annealing treatment under an oxygen atmosphere for removal of interfacial contamination and structural recovery. b-d) Cross-sectional ADF images and corresponding FFT pattern of the t-NNO after annealing at 660 °C, showing the elimination of the amorphous interfacial layer and formation of an atomically clean interface. e) ADF intensity profile showing a lattice spacing of 0.392 nm with 17 and 16 atomic layers in the top and bottom membranes, respectively. At the interface (red arrow), the amorphous contrast is fully eliminated, and the interfacial lattice spacing is ~0.28 nm. f) High-resolution chemical profile of Na $K$ (green) and Nb $L$ (purple) across the interface of the annealed t-NNO, showing the formation of an ordered –Nb–Na–Nb– sequence.

Following the treatment, the structural characteristics of the annealed t-NNO were examined. As shown in low-magnification ADF images (**Figure 3b**), the dark regions previously observed at the interface have almost completely disappeared. Atomic-resolution imaging and fast Fourier transform (FFT) pattern reveals that pseudo-cubic lattice spacing along out-of-plane ($d_{(001)pc}$) remains at 0.392 nm (**Figure 3c, d**), while the interface becomes atomically flat and sharp. The intensity profile extracted from the ADF image indicates that the top and bottom membranes consist of 17 and 16 atomic layers, respectively, corresponding to an increase of approximately six layers (~2.4 nm) in each membrane (**Figure 3e**) compared with the pristine state (**Figure S2**). This increase in thickness likely originates from interfacial recrystallization and the incorporation of amorphous material that became ordered during annealing, effectively restoring the perovskite lattice across the interface. In particular, the interatomic spacing at the annealed interface was measured to be ~0.28 nm, corresponding to an out-of-plane contraction of about 25% relative to the bulk lattice distance (**Figure 3e**).

Additional atomic-resolution imaging confirms a consistent result of interfacial recrystallization phenomenon, consistently demonstrating out-of-plane lattice contraction (**Figure S6**). Considering the intrinsic coupling between structural, electronic, and chemical degrees of freedom in complex oxides,[39] this lattice modification implies that the interface has experienced structural reconstruction rather than mere geometric reconfiguration. While the atomically flat interface indicates structural coherence, compositional and electronic structure analyses are necessary to verify whether genuine chemical bonding is established across the junction and elucidate modified interfacial environment.

## 2.3. Chemical bonding between annealed t-NNO membranes

To probe the interfacial chemistry at the atomic scale and verify the establishment of chemical bonding, atomic layer resolved EDX and EELS analyses were performed. As revealed by high-resolution EDX profile (**Figure 3f**), a sequential repetition of Na (A-site) and Nb (B-site) atomic columns was observed across the interface, confirming the formation of a periodic -($NbO_2$)-(NaO)-($NbO_2$)- stacking sequence. Moreover, the interfacial Na atomic position is slightly displaced toward Nb columns in the bottom membrane, a shift (**Figure 3f**, blue arrow) that aligns with the out-of-plane contraction identified in the ADF intensity profile. The continuous stacking of A-site and B-site layers, together with the lateral offsets, provides further evidence for the establishment of chemical bonding across the interface.

Electronic structure analysis using EELS fine-structure measurements further corroborates the presence of interfacial bonding. As indicated by the navy and red boxes in the ADF image (**Figure 4a**), Nb *M*- and O *K*-edge spectra were extracted from the interface and three positions below and above the interface. The Nb *M*-edge exhibits a characteristic doublet in both the $M_3$ and $M_2$ regions (**Figure 4b**), consistent with the octahedral crystal field splitting observed in $NbO_2$ or $Nb_2O_5$.[40,41] Since the $M_{2,3}$-edge arises from Nb *3p* core levels to unoccupied *4d* states, it alone cannot resolve subtle bonding variations. Thus, we focus on the

O *K*-edge energy-loss near edge structure (ELNES), which exhibits two distinct features: peak A at ~534 eV and peak B at ~538 eV with an energy difference of ~4 eV (**Figure 4c**). This splitting arises from crystal-field effects reflected in the O *K*-edge fine structure, corresponding to transitions into O *2p* states hybridized with Nb *4d* $t_{2g}$ and $e_g$ orbitals.[42] The spectra acquired from the bottom (spectra 1–3) and top (spectra 5–7) membranes exhibits typical NaNbO$_3$ spectral features,[42] in which the intensity of peak A was generally higher than that of peak B. In contrast, this trend alters at the interface, where the intensity of peak A decreases (**Figure 4c**, red arrow), resulting in a spectrum showing nearly identical magnitudes for peaks A and B.

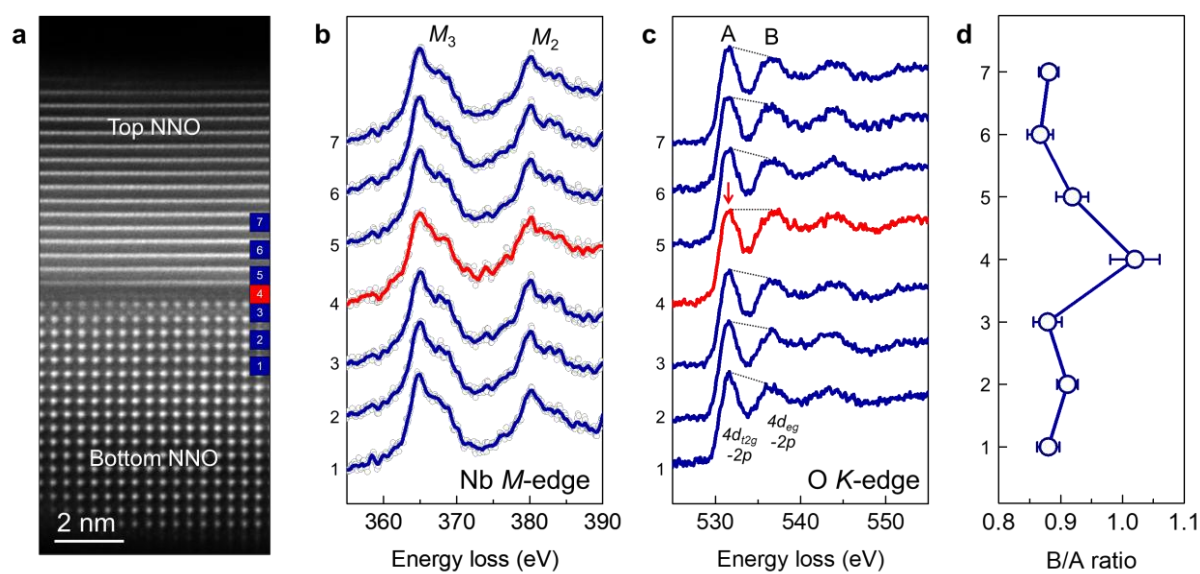

**Figure 4. Electronic structure analysis via electron energy loss spectroscopy (EELS).** a) ADF image of the annealed t-NNO, with navy and red boxes marking the positions selected for EELS acquisition. b, c) Nb *M*-edge spectra, and corresponding O *K*-edge ELNES, showing characteristic peak A (~534 eV) and peak B (~538 eV). The interfacial spectrum (red) exhibits suppression of peak A, yielding nearly equivalent intensities of peaks A and B. d) Quantitative analysis of the intensity ratio (B/A) for O *K*-edge spectra, showing an increased B/A value (~1.0) at the interface.

To quantify this change, we calculated the intensity ratio of peak B to peak A (B/A). Quantitative analysis reveals that the B/A ratio is approximately 0.9 for both the top and bottom membranes but increases to about 1.0 at the interface (**Figure 4d**). The change in the O *K*-edge intensity is likely associated with a modification of the local electronic environment at the strain-modulated interface. Although a slight Na deficiency could contribute to this variation, spectra 3 and 5, which correspond to regions exhibiting similar Na deficiency (**Figure 3f**), display intensity ratios that are comparable to those of the upper and lower layers (spectra 1, 2, 6, 7) but distinct from that of the interface spectrum. This finding suggests that the difference observed in the interfacial spectrum is more likely attributed to interfacial chemical bonding. Based on these spectral results (**Figures 3f and 4c**), the observed out-of-plane contraction (**Figure 3c, e**) can be predicted to originate from the formation of interfacial chemical accompanied by localized strain. Such behavior aligns with the mechanical characteristics of NNO, which exhibits relatively high ductility and compliance compared to other perovskite oxides.[43] Collectively, our imaging and spectroscopy results provide compelling evidence that the annealed t-NNO membranes form an atomically clean and electronically active interface supporting coherent interlayer bonding.

## 2.4. Local strain states in fully bonded t-NNO membranes

A key question for twisted oxide heterostructures, particularly in comparison with conventional epitaxial heterostructures, is whether strain is uniformly distributed across the interface, as local strain variations can strongly influence polarization configurations. To assess strain distribution within both membranes, atomic-resolution ADF images were acquired along the respective zone axes of the top and bottom membranes (**Figure 5a, b**). The peak pairs analysis (PPA) method was employed to extract and quantify the local strain states. Shear strain ($E_{xy}$) mapping reveals that the strain is not confined to the interface but extends throughout the top membrane (**Figure 5c**), indicating that interfacial modulation effectively propagates across the entire film. As a result, the top membrane exhibits a shear strain of $0.9 \pm 0.7\%$ (**Figure 5e**). In contrast, the bottom membrane shows a minimal strain of $0 \pm 0.4\%$ (**Figure 5d, f**), indicating that it is nearly relaxed.

To further verify the difference in strain behavior between the top and bottom membranes, we analyzed the in-plane ($E_{xx}$) and out-of-plane ($E_{yy}$) strain components. The results reveal pronounced strain variations in the top membrane, whereas the bottom membrane remains nearly strain-free (**Figure S7**). Additional $E_{xy}$ mapping confirms consistent strain distribution, corroborating the reproducibility of strain modulation across the top membrane (**Figure S8**). These asymmetric results suggest that the top membrane experiences greater strain modulation than the bottom one, likely associated with the stacking configuration. During annealing, the bottom membrane is mechanically constrained between the top membrane and the underlying Si substrate, such that the interfacial reconstruction does not significantly alter its relaxed strain state. In contrast, the top membrane retains the mechanical freedom to accommodate lattice adjustments and strain redistribution because of the interfacial recrystallization. Such mechanical compliance facilitates local lattice displacements and promotes their propagation of structural modifications throughout the entire top membrane. As a result, the t-NNO system relaxes substrate-induced mechanical constraints, thereby minimizing lattice mismatch effects and enabling the formation of a twisted bilayer structure without relying on epitaxial registry.

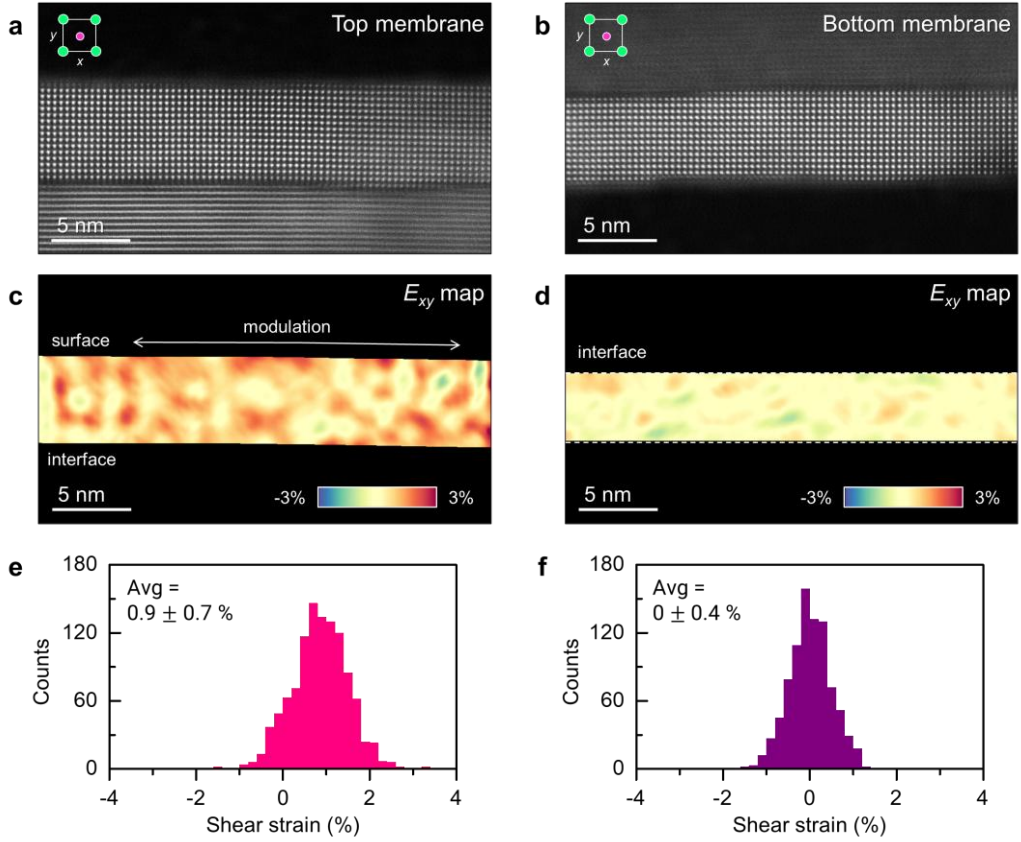

**Figure 5. Strain analysis of fully bonded t-NNO.** a, b) Atomic-resolution ADF images of top and bottom membranes in the annealed t-NNO. c, d) Corresponding shear strain ($E_{xy}$) maps. Notably, the top membrane exhibits strain propagation across the whole regions. e) Histogram for $E_{xy}$ extracted from the top membrane, showing positive shear with an average value of 0.9 ± 0.7%. f) Histogram for $E_{xy}$ extracted from the bottom membrane, showing nearly free states.

Such modulation facilitate controllable strain–coupling interactions distinct from those in substrate-clamped epitaxial films, with the resulting local distortions and moiré-induced strain patterns potentially influencing polarization switching,[28,44,45] flexoelectricity,[46,47] and energy-storage behavior.[20] Owing to the rich polymorphism, complex oxygen octahedral tilt patterns, and the coexisting ferroelectric and antiferroelectric orders[38] in NNO, its strong coupling to strain may give rise emerging strain-mediated topological polar textures, such as coupled ferroelectric and antiferroelectric configurations, that are distinct from those observed in twisted $BaTiO_3$ and $SrTiO_3$ heterostructures. Asymmetric strain fields extending throughout the top region may enable tunable ferroelectric and antiferroelectric phase transitions, potentially yielding reversible double hysteresis loops and offering prospects for high-energy-density, lead-free capacitor applications with improved environmental sustainability. In addition, the strain modulation in twisted moiré heterostructures perturbs local polarization and electronic structures, providing new opportunities for tailoring electromechanical and optoelectronic functionalities in oxide membranes. While this study focuses on elucidating the fundamental interfacial bonding mechanisms in twisted oxide membranes rather than demonstrating direct device applications, the insights gained here highlight the potential of twist engineering as a powerful route to tune interfacial structure, strain, and functionality in complex oxides.

## 3. Conclusion

In conclusion, we resolve a long-standing challenge in volatile-element-containing oxide twistronics by eliminating the interfacial amorphous "dead" layer that commonly forms in oxide membranes. Through identifying the origin of interfacial disorders in twisted freestanding $NaNbO_3$ and developing a practical oxidative annealing process, we achieve complete removal of the carbon-containing amorphous phase formed during aqueous processing. This treatment produces an atomically sharp, chemically bonded interface, with interlayer coupling directly confirmed by atomic-resolution imaging and spectroscopy. Furthermore, the observation of distinct strain states in the top and bottom membranes reveals asymmetric strain propagation due to the lattice softness of $NaNbO_3$, offering insights into engineering strain distributions in twisted oxides. These results overcome a critical bottleneck in oxide moiré fabrication and open new opportunities for designing chemically coherent, strain-tunable architectures hosting emergent ferroic and electronic phenomena.

## 4. Experimental Section/Methods

*Freestanding membranes synthesis:* Epitaxial heterostructures consisting of $NaNbO_3$ films with a thickness of 6.7 nm and 20 nm thick $La_{0.7}Sr_{0.3}MnO_3$ sacrificial layers were synthesized on (001)-oriented single-crystalline $SrTiO_3$ substrates via pulsed laser deposition. The $La_{0.7}Sr_{0.3}MnO_3$ sacrificial layer was synthesized in an oxygen partial pressure of 200 mTorr, at a growth temperature of 730 °C, a laser fluence of 1.7 J/cm$^2$, and a repetition rate of 3 Hz, using a 3.7 mm$^2$ imaged laser spot. The synthesis of $NaNbO_3$ films was conducted in an oxygen partial pressure of 210 mTorr, at a growth temperature of 660 °C, a laser fluence of 2.1 J/cm$^2$, and a repetition rate of 2 Hz, using a 4.58 mm$^2$ imaged laser spot. For membrane release and transfer, a 600 nm thick poly(methyl methacrylate) (PMMA) layer was spin-coated on top of the heterostructure and cured at 135 °C. The PMMA-coated sample was then immersed in a diluted etchant solution containing potassium iodide (5 g), hydrochloric acid (0.5 mL of 6 M concentration), and deionized water (50 mL) at room temperature. After 4 days of selective etching, the $La_{0.7}Sr_{0.3}MnO_3$ sacrificial layer was completely dissolved, allowing the PMMA/$NaNbO_3$ bilayer to detach from the $SrTiO_3$ substrate. The released membrane was transferred onto perforated silicon nitride ($SiN_x$) windows with circular holes on a silicon (Si) frame (Norcada Inc.). The PMMA support was subsequently removed by immersion in acetone at 80 °C followed by rinsing in isopropanol, yielding freestanding $NaNbO_3$ membranes suspended across the apertures. This process was repeated twice to prepare two separate freestanding membranes, which were then stacked with a relative angular offset to form twisted $NaNbO_3$ bilayers. Finally, the transferred twisted $NaNbO_3$ membranes were annealed under a dynamic oxygen flow at 660 °C for 2 hours.

*Structural characterizations:* High-resolution X-ray diffraction (HRXRD) measurements were carried out using a Rigaku SmartLab diffractometer to obtain θ–2θ line scans and rocking curves. Tapping mode atomic force microscopy was used to image the topography with an MFP-3D Origin+ AFM (Asylum Research) using a reflective aluminum-coated tip (BudgetSensors, Tap300Al-G, force constant $\approx 40$ N m$^{-1}$). Cross-sectional TEM specimens were fabricated using a focused ion beam (FIB) system (Helios NanoLab 450, FEI, Nanolab Technologies). Atomic-resolution imaging was conducted using annular dark-field scanning transmission electron microscopy (ADF-STEM) on a double Cs-corrected microscope (NeoARM200CF, JEOL, Japan) operated at 200 kV. The probe convergence and detector collection semi-angles were set to ~24 mrad and 68–270 mrad, respectively. Strain analysis was carried out using peak pairs analysis (PPA, HREM Research) applied to the ADF images. To suppress scanning distortion, a series of ADF frames were collected with a rapid acquisition time of 1 s and aligned drift correction. A Bragg filtering was applied by selecting the two primary reflections corresponding to the (100)$_{pc}$ and (001)$_{pc}$ lattice directions as reference vectors. Peak positions were subsequently identified in the averaged ADF images, from which the relative displacement fields ($u_x$, $u_y$) were computed. The strain tensors were defined as follows:

$$E_{xx} = \frac{\partial u_x}{\partial x} \quad E_{yy} = \frac{\partial u_y}{\partial y} \quad E_{xy} = \frac{1}{2}\left(\frac{\partial u_x}{\partial y} + \frac{\partial u_y}{\partial x}\right)$$

*Chemical characterizations:* Elemental mapping was performed by energy-dispersive X-ray spectroscopy (EDX) using a JED-2300T integrated into the ADF-STEM system. For EELS measurements, the core-loss EELS spectra of the Nb *M*- and O *K*-edge were obtained with a scan step of 0.2 Å across membranes. The energy dispersion was 0.03 eV/ch and the dwell time 0.05 s/pix with 10 passes using hybrid-pixel electron detector (Stela, Gatan, USA). Background subtraction was performed prior to core-loss signal extraction by fitting and removing the power-law background intensity. Min-max normalization between 0 and 1 was subsequently applied to the O *K*-edge spectrum over an energy-loss range of 525–555 eV and to the Nb *M*-edge spectrum over 350–395 eV. To evaluate the B/A peak intensity ratio, the local maxima corresponding to peak A (~534 eV) and peak B (~538 eV) were identified within the O *K* ELNES. To minimize the artifacts from spectral noise, intensity values were then extracted by averaging the signal over a 0.05 eV energy window centered at each peak position, and the B/A ratio was calculated from these averaged values. Error bars associated with the B/A ratio indicate local statistical variations, derived from the standard deviation of three spectral measurements collected along individual atomic rows of the same sample.

## Acknowledgements

This work was supported by the U.S. Department of Energy, Office of Basic Energy Sciences, Division of Materials Sciences and Engineering under contract FWP-ERKCS89. Microscopy technique development (Y.-H.K.) was partially supported by the U.S. BES, Early Career Research Program (KC040304-ERKCZ55). Experiments were performed partially at the Center for Nanophase Materials Sciences (CNMS), a U.S. DOE Office of Science User Facility at Oak Ridge National Laboratory (ORNL). R.G. acknowledges the support from the National Science Foundation (NSF) under award No. DMR-2442399. R.X. acknowledges support from the Army Research Office under award No. W911NF-25-1-0201. Y.-M.K. acknowledges the support of the National Research Foundation of Korea (NRF) grant (RS-2023-NR076943) funded by the Korean government.

## Conflict of Interest

The authors declare no conflict of interest.

## Author Contributions

Y-H.K., and R.G. contributed equally to this work. M.C., R.X. and Y-H.K. conceived this work. R.G., and R.X. prepared the twisted freestanding membranes and conducted XRD and AFM characterization. Y.-H.K., M.-H.J. and Y.-M.K. prepared TEM lamella specimens. Y.-H.K. conducted microscopy experiments. Y.-H.K. wrote the initial manuscript with contributions from R.X. and M.C.. All authors participated in editing the manuscript.

**Data Availability Statement**

The data that support the findings of this study are available from the corresponding author upon reasonable request.

**References**

1.  L. Du, M. R. Molas, Z. Huang, G. Zhang, F. Wang, Z. Sun, *Science* **2023**, *379*, eadg0014.
2.  M. Oudich, X. Kong, T. Zhang, C. Qiu, Y. Jing, *Nat. Mater.* **2024**, *23*, 1169.
3.  A. K. Geim, I. V. Grigorieva, *Nature* **2013**, *499*, 419.
4.  Y. Cao, V. Fatemi, S. Fang, K. Watanabe, T. Taniguchi, E. Kaxiras, P. Jarillo-Herrero, *Nature* **2018**, *556*, 43.
5.  M. Yankowitz, S. Chen, H. Polshyn, Y. Zhang, K. Watanabe, T. Taniguchi, D. Graf, A. F. Young, C. R. Dean, *Science* **2019**, *363*, 1059.
6.  M. Oh, K. P. Nuckolls, D. Wong, R. L. Lee, X. Liu, K. Watanabe, T. Taniguchi, A. Yazdani, *Nature* **2021**, *600*, 240.
7.  Y. Cao, V. Fatemi, A. Demir, S. Fang, S. L. Tomarken, J. Y. Luo, J. D. Sanchez-Yamagishi, K. Watanabe, T. Taniguchi, E. Kaxiras, R. C. Ashoori, P. Jarillo-Herrero, *Nature* **2018**, *556*, 80.
8.  K. P. Nuckolls, M. Oh, D. Wong, B. Lian, K. Watanabe, T. Taniguchi, B. A. Bernevig, A. Yazdani, *Nature* **2020**, *588*, 610.
9.  A. L. Sharpe, E. J. Fox, A. W. Barnard, J. Finney, K. Watanabe, T. Taniguchi, M. A. Kastner, D. Goldhaber-Gordon, *Science* **2019**, *365*, 605.
10. G. Chen, A. L. Sharpe, E. J. Fox, Y.-H. Zhang, S. Wang, L. Jiang, B. Lyu, H. Li, K. Watanabe, T. Taniguchi, Z. Shi, T. Senthil, D. Goldhaber-Gordon, Y. Zhang, F. Wang, *Nature* **2020**, *579*, 56.
11. M. Serlin, C. L. Tschirhart, H. Polshyn, Y. Zhang, J. Zhu, K. Watanabe, T. Taniguchi, L. Balents, A. F. Young, *Science* **2020**, *367*, 900.
12. N. P. Wilson, W. Yao, J. Shan, X. Xu, *Nature* **2021**, *599*, 383.
13. K. F. Mak, J. Shan, *Nat. Nanotechnol.* **2022**, *17*, 686.
14. S. S. Sunku, G. X. Ni, B. Y. Jiang, H. Yoo, A. Sternbach, A. S. McLeod, T. Stauber, L. Xiong, T. Taniguchi, K. Watanabe, P. Kim, M. M. Fogler, D. N. Basov, *Science* **2018**, *362*, 1153.
15. Y. Lee, X. Wei, Y. Yu, L. Bhatt, K. Lee, B. H. Goodge, S. P. Harvey, B. Y. Wang, D. A. Muller, L. F. Kourkoutis, W.-S. Lee, S. Raghu, H. Y. Hwang, *Nat. Synth* **2025**, *4*, 573.
16. N. Pryds, D.-S. Park, T. S. Jespersen, S. Yun, *APL Materials* **2024**, *12*, 010901.

17. S. Choo, S. Varshney, H. Liu, S. Sharma, R. D. James, B. Jalan, *Sci. Adv.* **2024**, *10*, eadq8561.
18. K. Eom, M. Yu, J. Seo, D. Yang, H. Lee, J.-W. Lee, P. Irvin, S. H. Oh, J. Levy, C.-B. Eom, *Sci. Adv.* **2021**, *7*, eabh1284.
19. D. Ji, S. Cai, T. R. Paudel, H. Sun, C. Zhang, L. Han, Y. Wei, Y. Zang, M. Gu, Y. Zhang, W. Gao, H. Huyan, W. Guo, D. Wu, Z. Gu, E. Y. Tsymbal, P. Wang, Y. Nie, X. Pan, *Nature* **2019**, *570*, 87.
20. G. Dong, S. Li, M. Yao, Z. Zhou, Y.-Q. Zhang, X. Han, Z. Luo, J. Yao, B. Peng, Z. Hu, H. Huang, T. Jia, J. Li, W. Ren, Z.-G. Ye, X. Ding, J. Sun, C.-W. Nan, L.-Q. Chen, J. Li, M. Liu, *Science* **2019**, *366*, 475.
21. Y. Li, C. Xiang, F. M. Chiabrera, S. Yun, H. Zhang, D. J. Kelly, R. T. Dahm, C. K. R. Kirchert, T. E. L. Cozannet, F. Trier, D. V. Christensen, T. J. Booth, S. B. Simonsen, S. Kadkhodazadeh, T. S. Jespersen, N. Pryds, *Advanced Materials* **2022**, *34*, 2203187.
22. R. Erlandsen, R. T. Dahm, F. Trier, M. Scuderi, E. Di Gennaro, A. Sambri, C. K. Reffeldt Kirchert, N. Pryds, F. M. Granozio, T. S. Jespersen, *Nano Lett.* **2022**, *22*, 4758.
23. J. Zhang, Y. Xie, K. Ji, X. Shen, *Applied Physics Letters* **2024**, *125*, 230503.
24. F. Gómez-Ortiz, L. Bastogne, S. Anand, M. Yu, X. He, P. Ghosez, *Phys. Rev. B* **2025**, *111*, L180104.
25. R. Mandal, S. Yun, K. Wurster, E. Dollekamp, J. N. Shondo, N. Pryds, *Nano Lett.* **2025**, *25*, 5541.
26. G. Sánchez-Santolino, V. Rouco, S. Puebla, H. Aramberri, V. Zamora, M. Cabero, F. A. Cuellar, C. Munuera, F. Mompean, M. Garcia-Hernandez, A. Castellanos-Gomez, J. Íñiguez, C. Leon, J. Santamaria, *Nature* **2024**, *626*, 529.
27. S. Zhang, L. Jin, Y. Lu, L. Zhang, J. Yang, Q. Zhao, D. Sun, J. J. P. Thompson, B. Yuan, K. Ma, Akriti, J. Y. Park, Y. H. Lee, Z. Wei, B. P. Finkenauer, D. D. Blach, S. Kumar, H. Peng, A. Mannodi-Kanakkithodi, Y. Yu, E. Malic, G. Lu, L. Dou, L. Huang, *Nat. Mater.* **2024**, *23*, 1222.
28. S. Lee, D. J. P. De Sousa, B. Jalan, T. Low, *Sci. Adv.* **2024**, *10*, eadq0293.
29. H. Wang, V. Harbola, Y. Wu, P. A. Van Aken, J. Mannhart, *Advanced Materials* **2024**, *36*, 2405065.
30. N. A. Shahed, K. Samanta, M. Elekhtiar, K. Huang, C.-B. Eom, M. S. Rzchowski, K. D. Belashchenko, E. Y. Tsymbal, *Phys. Rev. B* **2025**, *111*, 195420.
31. M. Schmidbauer, J. Maltitz, F. Stümpel, M. Hanke, C. Richter, J. Schwarzkopf, J. Martin, *Applied Physics Letters* **2025**, *126*, 101902.
32. H. KP, X. Wei, C.-H. Lee, D. Yoon, Y. Lee, K. J. Crust, Y.-T. Shao, R. Xu, J.-H. Kang, C. Liang, J. Park, H. Y. Hwang, D. A. Muller, **2025**, DOI 10.48550/ARXIV.2510.23042.
33. S. S. Hong, J. H. Yu, D. Lu, A. F. Marshall, Y. Hikita, Y. Cui, H. Y. Hwang, *Sci. Adv.* **2017**, *3*, eaao5173.

34. D. Lu, D. J. Baek, S. S. Hong, L. F. Kourkoutis, Y. Hikita, H. Y. Hwang, *Nature Mater* **2016**, *15*, 1255.
35. D. Pesquera, A. Fernández, E. Khestanova, L. W. Martin, *J. Phys.: Condens. Matter* **2022**, *34*, 383001.
36. H. Sha, Y. Zhang, Y. Ma, W. Li, W. Yang, J. Cui, Q. Li, H. Huang, R. Yu, *Nat Commun* **2024**, *15*, 10915.
37. R. Ghanbari, H. Kp, K. Patel, H. Zhou, T. Zhou, R. Liu, L. Wu, A. Khandelwal, K. J. Crust, S. Hazra, J. Carroll, C. J. G. Meyers, J. Wang, S. Prosandeev, H. Qiao, Y.-H. Kim, Y. Nabei, M. Chi, D. Sun, N. Balke, M. Holt, V. Gopalan, J. E. Spanier, D. A. Muller, L. Bellaiche, H. Y. Hwang, R. Xu, *Nat Commun* **2025**, *16*, 7766.
38. R. Xu, K. J. Crust, V. Harbola, R. Arras, K. Y. Patel, S. Prosandeev, H. Cao, Y. Shao, P. Behera, L. Caretta, W. J. Kim, A. Khandelwal, M. Acharya, M. M. Wang, Y. Liu, E. S. Barnard, A. Raja, L. W. Martin, X. W. Gu, H. Zhou, R. Ramesh, D. A. Muller, L. Bellaiche, H. Y. Hwang, *Advanced Materials* **2023**, *35*, 2210562.
39. S. Jesse, A. Kumar, T. M. Arruda, Y. Kim, S. V. Kalinin, F. Ciucci, *MRS Bull.* **2012**, *37*, 651.
40. D. Bach, H. Störmer, R. Schneider, D. Gerthsen, J. Verbeeck, *Microsc Microanal* **2006**, *12*, 416.
41. D. Bach, R. Schneider, D. Gerthsen, J. Verbeeck, W. Sigle, *Microsc Microanal* **2009**, *15*, 505.
42. Y. Yoneda, T. Kobayashi, T. Tsuji, D. Matsumura, Y. Saitoh, Y. Noguchi, *Jpn. J. Appl. Phys.* **2024**, *63*, 09SP12.
43. S. A. Khattak, S. M. Wabaidur, M. A. Islam, M. Husain, I. Ullah, S. Zulfiqar, G. Rooh, N. Rahman, M. S. Khan, G. Khan, T. Khan, B. Ghlamallah, *Sci Rep* **2022**, *12*, 21700.
44. I. Efe, B. Yan, M. Trassin, *Applied Physics Letters* **2024**, *125*, 150503.
45. H. Han, W. Li, Q. Zhang, S. Tang, Y. Wang, Z. Xu, Y. Liu, H. Chen, J. Gu, J. Wang, D. Yi, L. Gu, H. Huang, C. Nan, Q. Li, J. Ma, *Advanced Materials* **2024**, *36*, 2408400.
46. X. Liang, H. Dong, Y. Wang, Q. Ma, H. Shang, S. Hu, S. Shen, *Adv Funct Materials* **2024**, *34*, 2409906.
47. L. Wang, S. Liu, X. Feng, C. Zhang, L. Zhu, J. Zhai, Y. Qin, Z. L. Wang, *Nat. Nanotechnol.* **2020**, *15*, 661.